\documentclass[%
 reprint,
 amsmath,amssymb,
 aps,
]{revtex4-2}

\usepackage{graphicx}% Include figure files
\usepackage{dcolumn}% Align table columns on decimal point
\usepackage{bm}% bold math
\usepackage[dvipsnames]{xcolor}
\definecolor{Constructive}{HTML}{ff0000}
\definecolor{Strengthening}{HTML}{ffc020}
\definecolor{Weakening}{HTML}{00c000}
\definecolor{Destructive}{HTML}{0080ff}
\definecolor{Blue_g}{HTML}{0000FF}
\definecolor{Red_g}{HTML}{c0c0c0}

\usepackage{todonotes}

\begin{document}

%\preprint{APS/123-QED}

\title{Giant enhancement of transport driven by active fluctuations: impact of inertia}
% Force line breaks with \\

\author{K. Bia{\l}as}
\affiliation{Institute of Physics, University of Silesia, 41-500 Chorz{\'o}w, Poland}
%\author{J. {\L}uczka}
%\affiliation{Institute of Physics, University of Silesia, 41-500 Chorz{\'o}w, Poland}
\author{J. Spiechowicz}
\email{jakub.spiechowicz@us.edu.pl}
\affiliation{Institute of Physics, University of Silesia, 41-500 Chorz{\'o}w, Poland}

\begin{abstract}
Recently, a paradoxical effect has been demonstrated in which transport of a free Brownian particle driven by active fluctuations in the form of white Poisson shot noise can be significantly enhanced when it is additionally subjected to a periodic potential. This phenomenon can emerge in an overdamped system, but it may also be inertia-induced. Here, we considerably extend previous studies and comprehensively investigate the impact of inertia on the effect of free transport enhancement observed in the overdamped system. We detect that inertia can not only induce this phenomenon, but depending on a parameter regime, it may also strengthen, weaken, or even destroy it. We exemplify these different scenarios and explore the parameter space to identify the corresponding regions where they emerge. The variance of the active fluctuations amplitude distribution is a key determinant of the inertia influence on the effect of free transport amplification. Our results are relevant not only for microscopic physical systems but also for biological ones, such as, e.g., living cells, where fluctuations generated by metabolic activities are active by default.

%Recently, a new transport mechanism was discovered, where the transport of Brownian particles under the bias in the form of white Poisson shot noise can be enhanced compared to a free Brownian particle when it is subjected to the spatially symmetric periodic potential. It can be detected in both overdamped limit and when small nonzero inertia is added to the system, i.e., a strong damping regime. The latter regime is often overlooked in favor of simplified overdamped dynamics but may lead to completely different results. Here, we expand on our previous findings and focus on the influence of inertia on transport in comparison to the overdamped system in the whole mass spectrum.  We present how mass influences directed velocity and discuss when it is reduced or increased. We discuss under which conditions those behaviors emerge. Results presented in this study are relevant for transport in spatially periodic, both biological and artificial microscopic non-equilibrium systems, e.g., Josephson junctions or colloidal particles in optical potential.
\end{abstract}

%Use showkeys class option if keyword
\maketitle

%\tableofcontents
%\section{Introduction}

\section{Introduction}
A Brownian particle in a one-dimensional periodic potential is one of the fundamental models of statistical physics. It can describe a variety of currently important systems from Josephson junctions \cite{kautz,golubov,blackburn}, optical lattices \cite{lutz,barkai}, colloids in optical traps \cite{albay2018,paneru2021}, to transport in living cells \cite{bressloff}. Although it looks seemingly common and straightforward surprisingly it can capture a broad spectrum of unusual phenomena concerning both transport and diffusion such as ratchet effect \cite{hanggi2009, olson}, negative mobility \cite{machura2007, nagel2008, slapik2019, spiechowicz2019njp}, anomalous \cite{metzler2014,spiechowicz2016scirep,spiechowicz2017scirep} or giant \cite{reimann, lindner2016, bialas2020, spiechowicz2023entropy} diffusion.

In this work, we study another seemingly paradoxical transport effect. The average velocity of a free Brownian particle subjected to a constant force $f$ equals $\langle v \rangle = v_0 = f$. If this particle is additionally put into a periodic potential, then its directed velocity would naturally be hampered $\langle v \rangle < v_0$ \cite{risken} due to scattering on the potential barriers. However, it has recently been shown that this observation is no longer true if the particle is exposed to active fluctuations $\eta(t)$ with a bias $\langle \eta(t) \rangle = f$, rather than the constant force $f$. In such a setup, the average velocity of the particle in the periodic potential may be considerably enhanced in comparison to the free transport \cite{praca_w_PRE, mechanism}.

Active fluctuations are hallmarks of active matter systems \cite{shaebani, vrugt}. They are inherently nonequilibrium, which implies that they are not constrained by fundamental rules of physics like the fluctuation-dissipation theorem \cite{kubo, marconi} or detailed balance symmetry \cite{cates, gnesotto}. Active fluctuations may emerge due to either self-propulsion mechanism \cite{romanczuk, bechinger}, collisions of the system with active bath \cite{active1, active2, active3, active4, active5}, or a combination of these two factors. A prominent example are fluctuations generated by metabolic activity, which are active by default \cite{ariga_bio}. Understanding their role in living matter nowadays is a hot topic and a major challenge in physics \cite{active3, ezber, ariga}.

Active matter systems have been traditionally modeled using simplified overdamped dynamics. This fact poses several fundamental problems. The velocity exists therein only as an average observable. In such a model, fundamental quantities like velocity fluctuations and its higher moments, kinetic energy, cannot be defined \cite{jung,eichhorn}. Not to mention that it is always an approximation, since, in reality, there are no massless systems. That is why recently inertial active setups have gained interest in the community \cite{underdamped1, underdamped2, underdamped3, underdamped4, underdamped5}.

We follow this trend and study what is the impact of inertia on the enhancement of transport driven by active fluctuations outlined above. In our previous paper, we showed that inertia can induce this effect \cite{inertial}. Here, we considerably extend our predictions and offer a comprehensive view of this problem. We detect that, depending on the parameter regime, inertia can not only induce this paradoxical transport behavior but also destroy it. Strengthening or weakening of this effect via inertia is also possible. We not only exemplify each of these scenarios but also explore the space of parameters describing active fluctuations to identify the corresponding regions of their emergence. We identify the variance of the amplitude distribution as a key determinant of the inertia influence on the effect of free transport amplification.

The paper is organized as follows. In the next section, we introduce the model and details of the form of the active fluctuations as well as a quantity of interest. Then, in Sec. III we present our results by first inspecting the occurrence of free transport enhancement in the overdamped system and later studying the impact of inertia. Both general remarks and specific examples are discussed there. Finally, the last section provides a summary and final conclusions.

%This paper is split into four sections, in Sec. II, we introduce our model together with details of a form of stochastic force $\eta(t)$, scaling procedure of the Langevin equation describing the studied system and quantities of interest relevant to this study.

%Then in Sec. III, we briefly introduce the parameter regime where enhancement occurs in the overdamped dynamics and present an overview of how the addition of inertia to the system dynamics may affect transport compared to the overdamped system. The influence of mass may be negative, i.e., the average velocity is reduced, and enhancement either weakens or disappears. The opposite may also be true, i.e., the enhancement may emerge for system parameters, where transport was reduced compared to the free Brownian particle, or the magnitude of velocity may be higher than in the overdamped system.

%Next, we focus on a more detailed dependence of the average velocity on inertia and potential barrier height in each of those regimes and present some general properties of how transport in the steep periodic structure changes with the change of mass.

%We finish the Results section by describing the characteristics of active fluctuations necessary for each situation to appear and briefly introduce differences between overdamped and inertial dynamics in the studied system. Last but not least, Sec. IV provides a brief summary and final conclusions.

\section{Model}
Our system of interest is modeled by the dimensionless Langevin equation for an inertial Brownian particle of mass $m$ in a periodic potential $U(x)$ subjected to both active $\eta(t)$ and thermal $\xi(t)$ fluctuations, namely
\begin{equation}
    m\ddot{x}+ \gamma\dot{x} = -U'(x) + \eta(t) + \sqrt{2 D_T}\, \xi(t)
    \label{model}
\end{equation}
where $\gamma$ stands for the friction coefficient. The parameter $D_T$ determines the intensity of thermal noise, which is represented by a $\delta$-correlated Gaussian process of vanishing mean
\begin{equation}
	\langle \xi(t) \rangle = 0, \quad \langle \xi(t)\xi(s) \rangle = \delta(t-s).
\end{equation}
We refer the reader to Ref. \cite{inertial} for details of the scaling procedure, where the above dimensionless equation is obtained from a dimensional one. The following parameters are fixed
\begin{equation}
	\gamma = 1, \quad D_T = 0.01.
\end{equation}
We assume that the periodic potential has a simple spatially symmetric form
\begin{equation}
	U(x) = \varepsilon \sin{x}.
\end{equation}
with the barrier height $2\varepsilon$ and the period $L = 2\pi$. 

Active fluctuations $\eta(t)$ are represented by a sequence of $\delta$-shaped pulses with random amplitudes $z_i$ modeled as the Poissonian white shot noise \cite{hanggi1980,epl,spiechowicz2014pre,bialas2020,mechanism}
\begin{equation}
    \eta(t)=\sum_{i=1}^{n(t)}z_i\delta(t-t_i),
\end{equation}
where $t_i$ are the arrival times of the Poissonian counting process $n(t)$ \cite{feller1970}, i.e. the probability for occurrence of $k$ impulses in the time interval $[0,t]$ is
\begin{equation}
    Pr\{n(t)=k\}=\frac{(\lambda t)^k}{k!}e^{-\lambda t},
\end{equation}
parameter $\lambda$ determines how many $\delta$-pulses occur per unit of time on average and is called the mean spiking rate. Amplitudes $\{z_i\}$ are independent random variables drawn from a common probability distribution $\rho(z)$, which in our case is assumed to be the skew-normal density \cite{azz,rijal2022}. Such a choice can render the stochastic release of energy in the metabolic activity, such as, e.g., ATP hydrolysis or random collisions with the suspension of active particles. In this sense, this model is appropriate for both active particle self-propelling in a passive medium \cite{shaebani,romanczuk,bechinger} or a passive system immersed in an active bath \cite{active1,active2,active3,active4,active5}.

The skew-normal distribution $\rho(z)$ is usually defined in terms of three parameters, namely a location $\mu$ corresponding to the shift, a scale $\omega$ describing the spreading, and a shape parameter $\alpha$ affecting the asymmetry of the distribution. It has an analytic form, although it is not closed, namely
\begin{equation}
    \rho(z) = \frac{2}{\sqrt{2\pi \omega^2}}e^{-\frac{(z-\mu)^2}{2\omega^2}} \int_{-\infty}^{\alpha[(z-\mu)/\omega]} \frac{1}{\sqrt{2\pi}}e^{-\frac{s^2}{2}} ds.
\end{equation}
The quantities $\mu$, $\omega$, and $\alpha$ are not intuitive for physical applications. Thankfully, one can express them in terms of the statistical moments of the distribution $\rho(z)$, i.e., its mean $\zeta = \langle z_i\rangle$, variance $\sigma^2 = \langle (z_i-\zeta)^2\rangle$, and skewness $\chi = \langle (z_i-\zeta)^3\rangle/\sigma^3$ \cite{sp_generacja,sp_generacja2}
\begin{subequations}
\begin{align}
\begin{split}
\alpha&=\frac{\delta}{\sqrt{1-\delta^2}}, \\
\end{split}\\
\begin{split}
\omega&=\sqrt{\frac{\sigma^2}{1- 2\delta^2/\pi}}, \\
\end{split}\\
\begin{split}
\mu&=\zeta-\delta\sqrt{\frac{2\sigma^2}{\pi(1-2\delta^2/\pi)}},\\
\end{split}
\end{align}
\label{eq_S_def}
\end{subequations}
where $\delta$ is defined as
\begin{equation}
    \delta=\text{sgn}(\chi)\sqrt{\frac{|\chi|^{2/3}}{(2/\pi)\{[(4-\pi)/2]^{2/3}+|\chi|^{2/3}\}}}.
    \label{eq_S_delta}
\end{equation}
The nonequilibrium stochastic force $\eta(t)$ is  white noise with finite mean and covariance given by
\begin{subequations}
\begin{align}
\begin{split}
\langle\eta(t)\rangle&=\lambda \zeta, \\
\end{split}\\
\begin{split}
\langle\eta(t)\eta(s)\rangle-\langle\eta(t)\rangle\langle\eta(s)\rangle&=\lambda(\sigma^2+\zeta^2)\delta(t-s), \\
\end{split}
\end{align}
%\label{eq_S_def}
\end{subequations}
Moreover, we assume that active fluctuations are uncorrelated with thermal ones, namely
\begin{equation}
\langle\eta(t)\xi(s)\rangle=\langle \eta(t) \rangle \langle \xi(s) \rangle=0.
\end{equation} 

Unfortunately, equation (\ref{model}) cannot be solved in an analytic way. Therefore, to analyze our system of interest, we resort to precise numerical simulations using a Monte Carlo integration scheme \cite{euler} implemented on modern graphical processing units \cite{spiechowicz2015cpc}. The latter fact allowed us to speed up necessary calculations by several orders of magnitude as compared to traditional methods using central processing units.

In this paper, we aim to investigate the influence of inertia on the directed transport of a Brownian particle. For this reason, we will focus on the most fundamental quantity characterizing such a process, namely the long- -time average velocity
\begin{equation}
\langle v \rangle= \lim_{t\to\infty} \frac{\langle x(t) \rangle - \langle x(0) \rangle}{t}
\end{equation}
%\begin{equation}
%    \langle v\rangle=\lim_{t\to\infty}\frac{1}{t}\int_0^t \langle \dot{x}(\tau)\rangle d\tau
%\end{equation}
%in the overdamped limit, its expression has a simpler form
%\begin{equation}
%    \langle v\rangle=\lim_{t\to\infty}\frac{\langle x(t)-x(0)\rangle}{t},
%\end{equation}
where $\langle \cdot \rangle$ represents averaging over the ensemble of system trajectories and initial conditions. We will refer to the average velocity in the inertial system described by Eq. (\ref{model}) as $v_m$, in its overdamped counterpart for $m = 0$ as $v_{\gamma}$, and for the free Brownian particle ($U(x) = 0$) as $v_0$. Note that in the absence of the periodic potential all three quantities are equal, i.e. $v_0 = v_{\gamma} = v_m = \langle \eta(t) \rangle$.
\begin{figure}[t]
    \centering
    \includegraphics[width=0.9\linewidth]{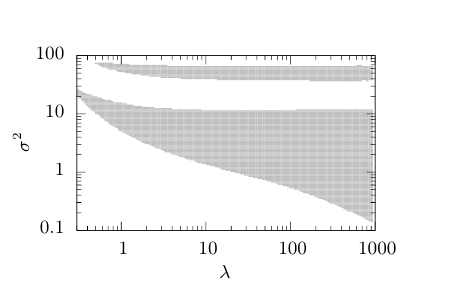}
    \caption{The average velocity $v_\gamma$ of an overdamped ($m = 0$) Brownian particle in a periodic potential $U(x)$ and driven by active fluctuations $\eta(t)$ as a function of their spiking rate $\lambda$ and variance $\sigma^2$. Other parameters are the mean bias $\langle \eta(t) \rangle = \lambda \zeta = 1$, skewness $\chi = 0.99$ and half of the potential barrier height $\varepsilon = 250$. The regions where the average velocity of the particle dwelling in the periodic potential is larger than for the free particle, i.e. $v_\gamma > v_0$, are marked with grey \textcolor{Red_g}{$\blacksquare$} color and the white background represents regions without the enhancement.}
   %Map of transport enhancement compared to free Brownian particle $\langle v\rangle>v_0$  as a function of mean spiking rate $\lambda$ and variance $\sigma^2$ for mean bias $\langle \eta(t)\rangle=\lambda \zeta=1$, skewness $\chi=0.99$, thermal fluctuation intensity $D_T=0.01$ and potential barrier height $\varepsilon=250$. Enhancement corresponds to \textcolor{Red_g}{$\blacksquare$} color.}
    \label{fig:0}
\end{figure}
\begin{figure}[t]
    \centering
    \includegraphics[width=0.9\linewidth]{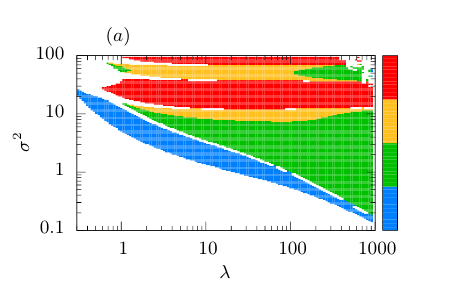}
    \includegraphics[width=0.9\linewidth]{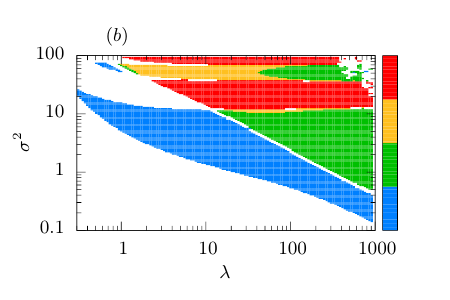}
    \includegraphics[width=0.9\linewidth]{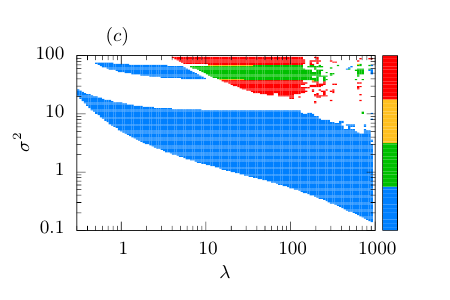}
    \caption{The average velocity $v_m$ of an inertial ($m \neq 0$) Brownian particle in a periodic potential $U(x)$ and driven by active fluctuations $\eta(t)$ as a function of their spiking rate $\lambda$ and variance $\sigma^2$. Other parameters are the mean bias $\langle \eta(t)\rangle=\lambda \zeta=1$, skewness $\chi=0.99$, and half of the potential barrier height $\varepsilon=250$. The parameter sets are marked with the corresponding color depending on the influence of the inertia on the average velocity: strengthening \textcolor{Strengthening}{$\blacksquare$} $v_m > v_\gamma > v_0$, weakening \textcolor{Weakening}{$\blacksquare$} $v_0 < v_m < v_\gamma$, destructive \textcolor{Destructive}{$\blacksquare$} $v_m < v_0 < v_\gamma$ or constructive (inertia-induced) \textcolor{Constructive}{$\blacksquare$} $v_\gamma < v_0 < v_m$. In panels (a), (b) and (c), the particle mass reads $m=0.01$, $m=0.1$ and $m=1$, respectively.}   
%Maps of enhancement regimes as a function of mean spiking rate $\lambda$ and variance $\sigma^2$ for mean bias $\langle \eta(t)\rangle=\lambda \zeta=1$, skewness $\chi=0.99$, thermal fluctuation intensity $D_T=0.01$ and potential barrier height $\varepsilon=250$. In panels (a), (b), and (c), specific regimes related to enhancement change in comparison to the overdamped limit (Constructive \textcolor{Constructive}{$\blacksquare$}, Strengthening \textcolor{Strengthening}{$\blacksquare$}, Weakening \textcolor{Weakening}{$\blacksquare$} and Destructive \textcolor{Destructive}{$\blacksquare$}) are shown for masses $m=0.01$, $m=0.1$ and $m=1$ in panels (a), (b) and (c) respectively.}
    \label{fig:1}
\end{figure}

\section{Results}
We start our analysis of influence of inertia on the giant enhancement of transport generated by active fluctuations by considering parameter regimes of nonequilibrium stochastic force $\eta(t)$ for which periodic potential can enormously boost free particle transport driven by active fluctuations $\eta(t)$ in the overdamped system $m = 0$ \cite{praca_w_PRE, mechanism}. The latter behavior is counterintuitive, as one would typically expect the transport of a free Brownian particle to be significantly slowed down due to the existence of the periodic barriers in the potential $U(x)$ \cite{risken}. However, for tailored parameter regimes, the average velocity $v_\gamma$ of the particle in the periodic structure $U(x)$ can be significantly larger than for the free particle $v_0$, i.e., $v_\gamma > v_0$ \cite{praca_w_PRE, mechanism}. We present them in Fig. \ref{fig:0}. There are two disjoint enhancement regions. A triangular one where the variance  $\sigma^2$ of amplitude distribution $\rho(z)$ is small, and a single $\delta$-impulse can transport the particle only over the nearest potential barrier and a thinner strip-like area for the larger variance.

The crucial reason of the free transport enhancement is the relaxation of the Brownian particle towards the potential minimum after the arrival of $\delta$-spike. Due to the asymmetric form of the amplitude distribution $\rho(z)$ relaxation in the direction of the average force $\langle \eta(t) \rangle$ is more likely than in the opposite one. This extra displacement, which is missing in the case of free particle, leads to the enhancement of the particle velocity when it is additionally subjected to the periodic potential. The transport amplification is optimized in the resonance regime when the mean time $\langle \tau_P \rangle = 1/\lambda$ between consecutive $\delta$-spikes is tuned to the average time $\langle \tau_R \rangle \propto 1/\varepsilon$ of the particle relaxation towards  $U(x) = \varepsilon \sin{x}$ minimum, i.e. $\langle \tau_P \rangle \approx \langle \tau_R \rangle$. In such a case the motion is synchronized, the next $\delta$-spike arrives when the relaxation process is completed and the particle fully exploited the periodic potential. The free transport enhancement emerges also in the regime of rare spikes, i.e. $\langle \tau_P \rangle \gg \langle \tau_R \rangle$, but its magnitude is smaller than in the resonance one.  Due to the detuning of $\langle \tau_P \rangle$ and $\langle \tau_R \rangle$ the particle needs to wait a considerable amount of time in the potential minimum for the arrival of the next $\delta$-spike and this fact hampers its transport velocity. Finally, when the $\delta$-pulses are frequent, i.e. $\langle \tau_P \rangle \ll \langle \tau_R \rangle$, there is no enhancement since then the particle is constantly agitated by active fluctuations and it does not have enough time to relax in the periodic potential.

In Fig. \ref{fig:1}, we present how inertia can affect this transport enhancement. In doing so we calculated the directed velocity $\langle v \rangle$ for a grid of 150x150 parameter regimes corresponding to the overdamped and inertial system and compared them. Four distinct scenarios may occur for a given parameter regime. If the above-mentioned effect is present already in the overdamped system, i.e. $v_\gamma > v_0$, it can be either strengthened $v_m > v_\gamma > v_0$, weakened $v_0 < v_m < v_\gamma$ or destructed $v_m < v_0 < v_\gamma$. On the other hand, inertia may have a constructive influence on the free transport enhancement, i.e., $v_\gamma < v_0 < v_m$. It means that this effect is induced purely by inertia \cite{inertial}. The above four options are marked with the corresponding color, see the Fig. \ref{fig:1} caption. 
%\textcolor{red}{The white background represents parameters not belonging to either regime or for which the identification of the regime with sufficient certainty was impossible due to noise, e.g. white boundary between blue and green regions.}

As inertia grows, the effect of free transport enhancement $v_\gamma > v_0$ of the overdamped Brownian particle observed for the triangular region in Fig. \ref{fig:0} is mostly either weakened or destructed. The only exception from this negative impact is the small area in panel (a) for tiny mass $m = 0.01$ for which the transport amplification is strengthened. The latter influence quickly disappears for larger inertia, see panels (b) and (c). On the other hand, a small dose of inertia can strengthen the phenomenon already observed in the overdamped system for the great variance $\sigma^2$ and moderate spiking rate $\lambda$, c.f. the thinner strip-like area in Fig. \ref{fig:0}. If the latter parameter is sufficiently large, this effect is weakened instead. The positive impact of inertia on the free transport amplification is visible mainly in the region between the two areas marked by the red color in Fig. \ref{fig:0}. When we compare the panels (a), (b), and (c) of Fig. \ref{fig:1} corresponding to masses $m=0.01$, $m=0.1$, and $m=1$, respectively, the reader can note that the area of negative impact (weakening or destructive) of inertia on the free transport enhancement expands as $m$ grows while at the same time the positive one (strengthened or induced) shrinks. It leads us to the conclusion that in the inertial system, the effect of free transport amplification is much less populated in the parameter space than in the overdamped one. In the following subsections, we aim to take a closer look at the selected examples of the four classes of the impact of inertia on the average velocity mentioned above.

\begin{figure}[t]
    \centering
    \includegraphics[width=0.9\linewidth]{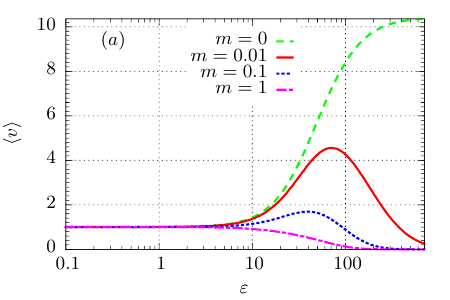}
    \includegraphics[width=0.9\linewidth]{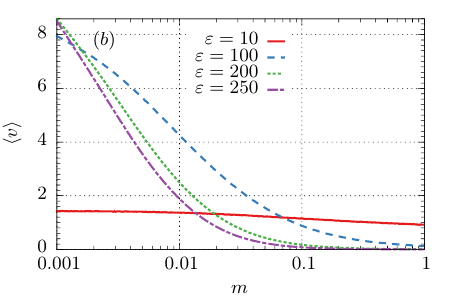}
    \caption{An illustration of the negative influence of inertia on the effect of free transport enhancement. The average velocity $\langle v \rangle$ of the Brownian particle driven by active fluctuations $\eta(t)$ as a function of the periodic potential $U(x)$ barrier height $\varepsilon$ for different values of mass $m$ is shown in panel (a). In plot (b), the same quantity is depicted versus $m$ and for different barrier heights $\varepsilon$. Other parameters read $\langle \eta(t)\rangle= \lambda \zeta=1$, variance $\sigma^2=2$, skewness $\chi=0.99$. In panel (a) the barrier height is $\varepsilon=250$ and in (b) the spiking rate is $\lambda=50$.}
    \label{fig:2}
\end{figure}
\subsection{Negative influence of inertia on the free transport enhancement}
Let us start with an illustration of the negative influence (weakening and destructive) of inertia on the effect of free transport amplification. It is presented in Fig. \ref{fig:2} where we depict the dependence of the average velocity $\langle v \rangle$ of the inertial Brownian particle in a periodic potential $U(x)$ and driven by active fluctuations $\eta(t)$ as a function of the barrier $\varepsilon$ and mass $m$. 

Panel (a) demonstrates one of the profound differences between the overdamped and inertial systems. It is seen for a large barrier height $\varepsilon$ for which the average velocity $\langle v \rangle$ in the presence of inertia $m \neq 0$ vanishes, whereas in the overdamped limit $m = 0$ it reaches the plateau. The reason for this behavior lies in the fact that in such a case, active fluctuations do not provide a sufficient portion of kinetic energy to cross the potential barrier. As a result, there exists an optimal barrier height $\varepsilon_o$ for which the free transport enhancement is maximized. Both $\varepsilon_o$ and $\langle v \rangle(\varepsilon_o)$ are larger for smaller inertia. One can also note that transport amplification ceases to emerge for a small barrier $\varepsilon$ as then the average velocity is equal to that of a free particle without the periodic potential, i.e. $\langle v \rangle = v_0$.
%\textcolor{red}{For small barrier heights $\varepsilon<\varepsilon_o$ and inertia $m\to0$, the results are qualitatively the same as in the overdamped limit, but not quantitatively, as the magnitude of the directed velocity is smaller in the inertial system.}

In panel (b), we show the influence of inertia $m$ on the average velocity $\langle v \rangle$ of the studied system. Regardless of the potential barrier height $\varepsilon$, this characteristic is a monotonically decreasing function of $m$. One can observe the existence of the critical mass $m_c$ for which $\langle v \rangle(m_c) = v_0 = 1$. It means that for $m > m_c$ the effect of free transport enhancement is destructed. The larger the barrier height $\varepsilon$, the smaller the critical mass $m_c$. If $m < m_c$, the effect of transport amplification is weakened as compared to the overdamped situation. We observe here a qualitative agreement between the overdamped $m = 0$ and strongly damped $m \to 0$ regimes, but not quantitative as the magnitude of the directed velocity $\langle v \rangle$ for small but finite $m$ is notably smaller than in the overdamped limit.
%but the average velocity of the inertial particle $\langle v \rangle$ is smaller than in the corresponding overdamped system $v_0 = 1 < \langle v \rangle < v_\gamma = \langle v \rangle (m \to 0)$ the effect of free transport amplification is weakened. 

\begin{figure}[t]
    \centering
    \includegraphics[width=0.9\linewidth]{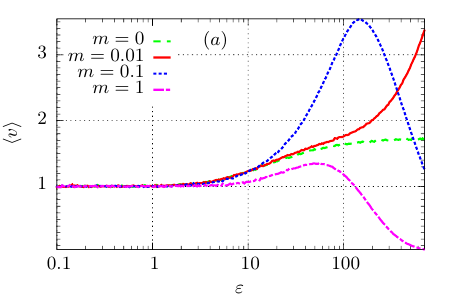}
    \includegraphics[width=0.9\linewidth]{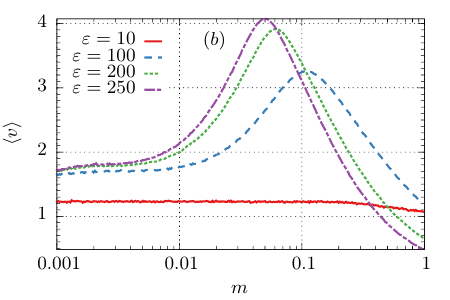}
    \caption{An illustration of the positive influence (strengthening) of inertia on the effect of free transport enhancement. The average velocity $\langle v \rangle$ of the system as a function of the barrier height $\varepsilon$ for different masses $m$ (panel (a)) and versus mass $m$ and various barrier heights (panel (b)). Other parameter read $\langle \eta(t) \rangle=\lambda \zeta=1$, variance $\sigma^2=50$, skewness $\chi=0.99$. In panel (a), the barrier height is $\varepsilon=250$ and in panel (b), the spiking rate is $\lambda=5$.}
    \label{fig:4}
\end{figure}
\subsection{Positive influence of inertia on the free transport enhancement}
\subsubsection{Strengthening}

In Fig. \ref{fig:4} we show the characteristics $\langle v \rangle(\varepsilon)$ and $\langle v \rangle(m)$ for different parameter regimes in which the influence of inertia on the studied effect is positive, i.e., it is strengthened. The dependence $\langle v \rangle(\varepsilon)$ displays similar qualitative features as in the previous example, i.e., the plateau in the overdamped limit and the non-monotonic behavior with the distinct maximum in the inertial system. However, since for low to moderate inertia $m$ the average velocity $\langle v \rangle$ is larger than in the overdamped situation ($m = 0$) and it vanishes for $m \to \infty$ we observe the transition from the positive (strengthening) to the negative (weakening and later destructive) influence of inertia, see e.g. the curve for $m = 0.1$.

The dependence $\langle v \rangle(m)$ shown in panel (b) is qualitatively different from the one corresponding to the negative influence of inertia. In particular, instead of the monotonic decrease of the average velocity $\langle v \rangle$ observed in Fig. \ref{fig:2}(b), there exists an optimal mass $m_o$ for which the transport is maximized. Notably, $\langle v \rangle (m_o)$ is much greater than in the overdamped system. The optimal inertia $m_o$ is smaller for larger barrier heights $\varepsilon$ and might disappear when the periodic potential is not steep enough, e.g., for $\varepsilon=10$. In this parameter regime, we can observe how transport in the inertial system tends to the one corresponding to the overdamped limit when mass vanishes $m \to 0$, as the average velocity $\langle v \rangle$ hardly changes when $m < 0.005$. It means that we observe both qualitative and quantitative agreement between the overdamped $m = 0$ and strongly damped $m \to 0$ regimes.
%\textcolor{red}{It is one of the limited cases, where the overdamped approximation leads to the same results as the inertial dynamics quantitatively.}

\begin{figure}[t]
    \centering
    \includegraphics[width=0.9\linewidth]{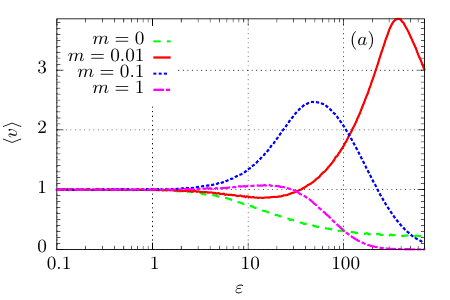}
    \includegraphics[width=0.9\linewidth]{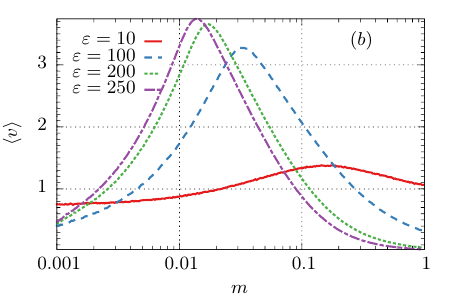}
    \caption{An illustration of the positive influence (constructive) of inertia on the effect of free transport enhancement. The average velocity $\langle v \rangle$ of the system as a function of the barrier height $\varepsilon$ for different masses $m$ (panel (a)) and versus mass $m$ and various barrier heights (panel (b)). Other parameter read $\langle \eta(t)\rangle=\lambda \zeta=1$, variance is $\sigma^2=20$, skewness $\chi=0.99$. In panel (a), the barrier height is $\varepsilon=250$ and in panel (b), the spiking rate is $\lambda=5$.}
    \label{fig:5}
\end{figure}
\subsubsection{Constructive}
Response of the system in the constructive regime in which the effect of free transport enhancement is induced by inertia is similar to the previous example, where strengthening is observed; however, there are some subtle differences. A new feature emerges in the dependence of the average velocity $\langle v \rangle$ on the barrier height $\varepsilon$ for small mass $m$ as presented in Fig. \ref{fig:5} (a), i.e., there exists a local minimum $\varepsilon_{min}\approx 15$ for $m=0.01$. Moreover, there is also an optimal portion of inertia for which the directed transport is maximized. The latter is smaller when the periodic potential has a larger barrier. The main difference is in the overdamped limit of vanishing inertia $m \to 0$. In this case, for strengthening regime $\langle v \rangle > v_0$, whereas here $\langle v \rangle < v_0$, i.e., the effect of free transport enhancement does not emerge. Consequently, the strongly damped $m \to 0$ regime is both qualitatively and quantitatively different than the overdamped limit $m = 0$.
%\textcolor{red}{This regime is both quantitatively and qualitatively different than the overdamped limit as small inertia $m<0.01$ leads to completely different directed transport and dependence on barrier height $\varepsilon$.}

\begin{figure}[t]
    \centering
    \includegraphics[width=0.9\linewidth]{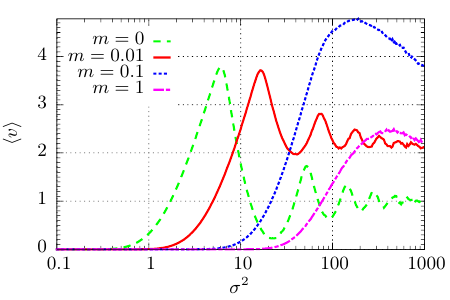}
    \caption{The average velocity $\langle v\rangle$ of the particle as a function of variance $\sigma^2$ for different values of its mass $m$. Other parameters are $\langle \eta(t)\rangle= \lambda \zeta=1$, the barrier height is $\varepsilon=250$, the spiking rate $\lambda=5$ and the skewness $\chi=0.99$.}
    \label{fig:3}
\end{figure}
%\subsection{Conditions for emergence of the presented regimes}
\subsection{Variance as a determinant of inertia influence}
All previously presented scenarios of influence of inertia on free transport enhancement emerge under a specific choice of the parameter regime, as depicted in Fig. \ref{fig:1}. An immediate conclusion from inspecting these results is that for the fixed spiking rate $\lambda$ we may observe all scenarios of inertia impact on the transport amplification, i.e. strengthening, weakening, constructive or destructive, depending on the value of the amplitude variance $\sigma^2$. Therefore, the latter parameter plays a role of a key determinant of the inertia influence.

In Fig. \ref{fig:3} we study this issue in detail. We present there the average velocity $\langle v \rangle$ of the particle as a function of the amplitude variance $\sigma^2$ for different inertia $m$. In the overdamped limit $m = 0$, we observe characteristic oscillations around the value corresponding to the free transport, i.e., $\langle v \rangle = v_0 = 1$. For the inertial system $ m\neq 0$, their reminiscence is still visible, but the directed transport $\langle  v\rangle$ no longer oscillates around $v_0$. The rule is, however, that for small variances $\sigma^2$ the amplitudes of active fluctuations are too small to take the particle over the potential barrier and therefore the average velocity $\langle v \rangle$ vanishes. There is a critical value of $\sigma^2$ for which the directed transport emerges. The latter is the greater, the larger the inertia is. 

The variance $\sigma^2$ determines how mass can affect the effect of free transport enhancement. In the overdamped limit $m = 0$, there exists a characteristic value $\sigma_o^2 \approx 5$ corresponding to the first maximum of the average velocity $\langle v \rangle$. If $\sigma^2 < \sigma_o^2$, then the directed transport can be only reduced in the inertial system with $m \neq 0$. For $\sigma^2 > \sigma_o^2$, all scenarios of positive and negative influence of inertia on the free transport amplification may be observed, depending on whether the average velocity $v_\gamma$ in the overdamped limit is smaller or larger than the value $v_0$ corresponding to the free particle. However, one can conclude that for the strong damping regime of minute mass $m \ll 1$, the phenomenon of our interest is typically either induced (constructed) or strengthened by inertia.

The reason why a larger variance $\sigma^2$ of active fluctuations $\eta(t)$ amplitudes $z_i$ is required for transport to emerge as mass $m$ is increased lies in a significant difference between the overdamped and inertial dynamics. Let us now discuss it in more detail. In doing so, we focus on the problem of how a single $\delta$-impulse acts on the Brownian particle residing in the position corresponding to the periodic potential $U(x)$ minimum. 

In the overdamped system described by the first-order Langevin equation, active fluctuations $\eta(t)$ operate in the particle coordinate space $\{x\}$. As a consequence, a single $\delta$-impulse with the amplitude $z$ moves the particle directly by the displacement $\Delta x = z$. It means that in principle it can take the particle over an infinite potential barrier if the amplitude is larger than the critical one $z > z_c = L/2 = \pi$ corresponding to the distance between the potential $U(x)$ minimum and the nearest maximum. Therefore, in the overdamped limit, there is an inherent coupling between the critical amplitude $z_c$ and the spatial geometry $L$ (period) of the potential $U(x)$.

On the other hand, in inertial dynamics $m \neq 0$ described by the Langevin equation of the second order, active fluctuations $\eta(t)$ act in the particle velocity (momentum) space $\{v\}$. A single $\delta$-impulse with the amplitude $z$ transfers the momentum $m \Delta v = z$ to the particle. It means that the particle must accumulate a sufficient portion of kinetic energy to overcome the potential barrier $2 \varepsilon$. In the Hamiltonian regime of non-dissipative dynamics with $\gamma \dot{x} = 0$, it can be calculated from the law of conservation of energy $mv_c^2/2 = 2\varepsilon$, i.e., the critical amplitude reads $z_c = 2 \sqrt{\varepsilon m}$. In such a case, there is a coupling between $z_c$ and the energetic barrier $\varepsilon$ of the potential $U(x)$ rather than its spatial geometry $L$. For the dissipative inertial dynamics such as Eq. (\ref{model}) the critical amplitude must be greater $z_c > 2 \sqrt{\varepsilon m}$ due to the damping which dissipates the energy during the uphill motion of the particle towards the top of the energetic barrier. 

The above discussion shows that the particle inertia $m$ affects the value of the critical amplitude $z_c$, which is in turn related to the probabilities to cross the potential barrier in both positive and negative directions. The balance between the latter quantities is directly connected with the average velocity $\langle v \rangle$ of the particle \cite{inertial}. On the other hand, the variance $\sigma^2$ of active fluctuations amplitudes $z_i$ shapes their statistics $\rho(z)$. Therefore, $\sigma^2$ also impacts the probabilities of crossing the potential barrier in both directions. The combination of these two contributions is responsible for the emergence of various regimes of the influence of inertia on the free transport enhancement effect, as shown in Fig. \ref{fig:3}.

%On the other hand, in inertial dynamics, an active fluctuation with amplitude $z$ represents an increase in the particle's velocity $\Delta v=z/m$. The damping term quickly reduces excess velocity. Note that the damping time is proportional to the inertia $\tau_D \propto m$. Therefore, if $m\to0$, we recover instantaneous displacement from the overdamped system.

%\begin{figure}[t]
%    \centering
%    \includegraphics[width=0.9\linewidth]{fig6_lam_5.pdf}
%    \caption{Average velocity $\langle v\rangle$ as a function of mass $m$ for different values of variance $\sigma^2$. Mean bias $\langle \eta(t)\rangle=\lambda \zeta=1$, the barrier height is $\varepsilon=250$, mean spiking rate $\lambda=5$, skewness $\chi=0.99$ and the temperature is $D_T=0.01$.}
%    \label{fig:6}
%\end{figure}
\section{Conclusions}
In summary, in this work, we studied the transport of an inertial Brownian particle dwelling in a spatially periodic symmetric potential and driven by active fluctuations in the form of white Poisson shot noise. We focused on the impact of the inertia on the effect of free transport enhancement, in which the average velocity of the system can radically exceed the value characteristic for the free particle driven by active fluctuations when it is additionally placed in the periodic potential.

Our system of reference was an overdamped counterpart of the studied setup. We revealed that inertia may have both a negative and a positive influence on the effect of free transport amplification emerging in the overdamped system. It can destruct this phenomenon or weaken the enhancement. On the other hand, there are cases when this factor can strengthen the amplification or even induce it. The latter situation means that the effect emerges in the inertial system, but it ceases to exist in its overdamped analog. We not only exemplified the above different scenarios but also explored the space of parameters describing active fluctuations to identify the corresponding regions where they emerge. It leads us to the conclusion that in the inertial setup, the studied phenomenon is much less populated than it is in the overdamped one. Moreover, the variance of the amplitude distribution is a key determinant of the inertia influence on the effect of free transport enhancement.

This work pushes much further our current understanding of the role of inertia in systems driven by active fluctuations. Since in reality there are no strictly overdamped setups with zero mass but rather strongly damped ones with small but nonzero inertia, our findings can be corroborated in a larger spectrum of readily available and tunable experimental setups consisting of, e.g., Josephson junctions \cite{kautz} or colloidal particles in optical traps \cite{albay2018, paneru2021}. The latter can be realized by using the optical tweezers and combining them with an ultrafast feedback control technique \cite{albay2018}. For this purpose a colloid should be suspended in a liquid and subjected to the harmonic potential generated by the optical tweezers. Then a high precision measurement of the particle position in the optical tweezers needs to be done. Next, the feedback force necessary for the generation of the effective potential of any desired shape is calculated. It is finally applied to the particle in the form of optical force via the ultrafast modulation of the trap center. The same technique can be exploited for the generation of the white Poissonian noise.

\section*{Acknowledgments}
This work has been supported by Grant NCN No. 2024/54/E/ST3/00257.

\section*{Data availability statement}
The data supporting this study findings are available from the corresponding author upon request.

%\nocite{*}

\end{document}